\documentclass[twocolumn,footinbib,aps,prl,floats,floatfix,amsmath,amssymb,secnumarabic]{revtex4-2}

\usepackage{float}
\usepackage{amsfonts}
\usepackage{amsmath}
\usepackage{slashed}
\usepackage{mathrsfs}
\usepackage{bbm}
\usepackage[scr=rsfs,cal=boondox]{mathalfa}
\usepackage{amssymb}
\usepackage{graphicx}
\usepackage{makeidx}
\usepackage{cancel}
\usepackage{dsfont}
\usepackage{latexsym}
\usepackage{mathtools}
\usepackage[dvipsnames]{xcolor}
\usepackage{float}
\usepackage{multirow}
\usepackage{xurl,hyperref}
\usepackage[compat=1.1.0]{tikz-feynman}
\usepackage{enumitem}
\usepackage{soul}
\hypersetup{colorlinks=true,citecolor=blue,linkcolor=magenta,urlcolor=NavyBlue}
\usepackage[utf8]{inputenc}
\usepackage{ragged2e}
\usepackage{natbib}
\usepackage{ulem}

\def\ba{\begin{array}}       
	\def\ea{\end{array}}

\def\beq{\begin{eqnarray}}
\def\eeq{\end{eqnarray}}
\begin{document}

\title{Is the Light Neutralino Dark Matter \\ Still Viable in the MSSM After the LZ-2024 Results?}

\author{Zhiyang Bao$^{a,b}$, Subhadip Bisal$^{a}$, and Junjie Cao$^{a,b}$}

\affiliation{$^a$ School of Physics, Zhengzhou University, Zhengzhou 450000, China\\
	$^b$ School of Physics, Henan Normal University, Xinxiang 453007, China}

\email{baozhiyang@stu.htu.edu.cn}
\email{subhadipbisal6@gmail.com (Corresponding author)}
\email{junjiec@alumni.itp.ac.cn (Corresponding author)}



\begin{abstract}
In the Minimal Supersymmetric Standard Model with heavy sfermions, a sub-hundred-GeV Bino-like 
neutralino can thermally achieve the observed dark matter abundance 
only through $h$- or $Z$-resonant annihilation. Since the couplings 
governing these resonances simultaneously determine the spin-independent and 
spin-dependent scattering off nucleons, reproducing the relic density 
imposes irreducible lower envelopes on the scattering cross sections. 
We evaluate these bounds including the complete one-loop
corrections to these vertices, which enhance the direct-detection rates by up to
$50\%$ and $30\%$, respectively, adopt the FLAG-2024 nucleon matrix
elements with their most adverse $3\sigma$ shifts, and confront the
predictions with a likelihood-based recast of the LZ data. 
We find that the \textbf{LZ-2024} results exclude both resonant 
scenarios at conservative one-sided significances exceeding $5.9\sigma$ and 
$4.1\sigma$, respectively, irrespective of whether the neutralino constitutes all or only part of
the dark matter, and independently of LHC electroweakino searches. 
\end{abstract}

\maketitle 

\textit{Introduction.}--- Identifying the particle nature of dark matter (DM) remains a central problem in particle physics and cosmology~\cite{Kolb:1990vq, Jungman:1995df, Bertone:2004pz}. Weakly interacting massive particles
(WIMPs), particularly those with mass below TeV scale, are among the most extensively studied candidates, since their
thermal freeze-out in the early Universe reproduces the observed relic
abundance, $\Omega h^2 = 0.120 \pm 0.001$~\cite{Hinshaw_2013, Planck:2018vyg}, largely independently of the initial conditions of cosmological evolution. In the
Minimal Supersymmetric Standard Model (MSSM) with $R$-parity conservation~\cite{Haber:1984rc,Gunion:1984yn,Djouadi:2005gj},
the lightest neutralino $\widetilde\chi_1^0$ provides a prototypical
realization of this mechanism~\cite{Jungman:1995df}. 

If sfermions are significantly heavier than the electroweak (EW) scale, 
as indicated by collider searches for supersymmetry~\cite{ATLAS:2024exu}, the minimal 
viable options become highly restrictive~\cite{Kowalska:2018toh}. A Higgsino- or Wino-like 
$\widetilde{\chi}^0_1$ below the TeV scale annihilates too efficiently 
and usually underproduces the relic density, whereas an almost pure Bino,
whose annihilation proceeds only through suppressed channels, generically
overcloses the Universe. A sub-100~GeV Bino-like $\widetilde{\chi}^0_1$ can
nevertheless reproduce the measured abundance through a small but essential
Higgsino admixture, provided its mass lies close to $m_Z/2$ or $m_h/2$, where its
annihilation is resonantly enhanced by $s$-channel $Z$- or Higgs-boson exchange~\cite{Drees:1992am, Pierce:2004mk, Grothaus:2012js, Han:2013gba, Cao:2013mqa, Han:2014sya, Calibbi:2014lga, Chakraborti:2015mra, Hamaguchi:2015rxa, Barman:2022jdg}.
Barring coannihilation configurations, these two resonance regions constitute the
only remaining thermal production mechanism for a sub-hundred-GeV
$\widetilde{\chi}^0_1$ in the MSSM with decoupled sfermions, and they are the subject
of this Letter.

Our focus is that this resonant mechanism has direct implications for DM searches. The
annihilation rate and the DM--nucleon scattering cross sections are governed by
the same $\widetilde{\chi}^0_1\widetilde{\chi}^0_1 h$ and
$\widetilde{\chi}^0_1\widetilde{\chi}^0_1 Z$ couplings, denoted by $C_h$ and $C_Z$, respectively.
Both arise from the Higgsino admixture of the predominantly Bino-like
$\widetilde{\chi}^0_1$ and are therefore suppressed as the Higgsino mass parameter
$\mu$ increases, as shown in Eq.~\eqref{eq:LOcouplings} of the Supplemental Material. In
the $h$ funnel, $C_h$ controls both the resonant annihilation and the
Higgs-mediated spin-independent (SI) cross section $\sigma^{\rm SI}$; likewise,
in the $Z$ funnel, $C_Z$ governs both the resonant annihilation and the
$Z$-mediated spin-dependent (SD) cross section $\sigma^{\rm SD}$. A resonance
does not remove this correlation; it merely allows the observed abundance to be
obtained with smaller couplings, and hence weaker scattering rates. This
reduction is bounded, however, since the finite mediator width and the thermal
smearing of the initial-state kinematics prevent the thermally averaged
annihilation cross section from becoming arbitrarily large~\cite{Griest:1990kh,Arcadi:2017kky}. Requiring
$\widetilde{\chi}^0_1$ to reproduce a specified relic abundance therefore imposes a
lower envelope on $\sigma^{\rm SI}$ in the $h$ funnel and on $\sigma^{\rm SD}$ in
the $Z$ funnel; both envelopes decrease with increasing relic abundance 
(or equivalently, increasing $\mu$) but remain finite.
The steadily improving sensitivity of the {\sc Lux-Zeplin} (LZ) experiment~\cite{LZ:2022lsv, LZ:2024zvo, LZ:2026axp}
can thus probe not only generic Bino--Higgsino mixtures but also the resonant
regions with the smallest scattering rates.

A robust assessment of these regions requires a consistent treatment of
theoretical and experimental uncertainties. We compute the next-to-leading-order
(NLO) supersymmetric corrections to the $\widetilde{\chi}^0_1\widetilde{\chi}^0_1 h$ and
$\widetilde{\chi}^0_1\widetilde{\chi}^0_1 Z$ vertices and propagate them to
both $\Omega h^2$ and $\sigma^{\rm SI,SD}$, thereby quantifying the residual EW uncertainty from missing higher orders. We further include the
uncertainties of the relevant nucleon matrix elements and analyze the LZ data
with a likelihood-based statistical treatment rather than a direct comparison
with the published limit curves. Even with these conservative refinements, we
find that \textbf{LZ-2024} excludes both the $h$- and the $Z$-funnel regions at more than
$4\sigma$. This exclusion holds whether $\widetilde{\chi}^0_1$ accounts for the total
DM abundance or only a subdominant component, and it is established independently
of LHC electroweakino (EWino) search constraints. Details of the involved calculations 
are given in the Supplemental Material.

\textit{DM--nucleon scattering.}---
In the non-relativistic limit, the effective Lagrangian governing the interactions of $\widetilde{\chi}_1^0$ with light quarks and gluons reads~\cite{Jungman:1995df, Bertone:2004pz}:
\begin{align}
\mathcal{L}^{\rm eff}=&\sum_{q=u,d,s}\left[\eta_q\overline{\widetilde{\chi}^0_1}\gamma^\mu\gamma_5\widetilde{\chi}^0_1\bar{q}\gamma_\mu\gamma_5 q + \lambda_qm_q\overline{\widetilde{\chi}_1^0}\widetilde{\chi}^0_1 \bar{q}q\right] \nonumber\\ 
 &+  \lambda_G\overline{\widetilde{\chi}^0_1}\widetilde{\chi}^0_1G^a_{\mu\nu}G^{a\,\mu\nu} + \dots~, 
\label{eq:L_eff}
\end{align}
where the axial-vector operator generates SD scattering, while the scalar quark and gluon operators 
contribute to SI scattering. The ellipsis denotes quark and
gluon twist-2 operators, which affect only subdominantly the SI scattering~\cite{Drees:1993bu}; 
their leading order (LO) contributions are included in \texttt{micrOMEGAs}~\cite{Belanger:2001fz, Belanger:2006is, Belanger:2008sj, Belanger:2013oya, Alguero:2023zol}.

At the nucleon level, the SI and SD cross sections of $\widetilde\chi^0_1$ off a nucleon $N\in\{p,n\}$ are
\begin{equation}
  \sigma^{\rm SI}_{N}=\frac{4}{\pi}\,\mu_{\widetilde{\chi}_1^0 N}^{2}\,f_{N}^{2},
  \qquad
  \sigma^{\rm SD}_{N}=\frac{12}{\pi}\,\mu_{\widetilde{\chi}_1^0 N}^{2}\,g_{N}^{2},
  \label{eq:sigma-def}
\end{equation}
with $\mu_{\widetilde{\chi}_1^0 N}=m_{\widetilde{\chi}_1^0}m_{N}/(m_{\widetilde{\chi}_1^0}+m_{N})$ the reduced mass and 
\begin{equation}
  \frac{f_{N}}{m_{N}}=\sum_{q=u,d,s}\lambda_{q}f^{N}_{q}
  -\frac{8\pi}{9\alpha_{s}}\lambda_{G}f^{N}_{G},
  \quad
  g_{N}=\sum_{q=u,d,s}\eta_{q}\,\Delta q_{N}.  \nonumber 
\end{equation}
Here $f^{N}_{q}=\langle N|m_{q}\bar qq|N\rangle/m_{N}$,
$f^{N}_{G}=1-\sum_{q}f^{N}_{q}$, and the quark spin fractions $\Delta q_N$ are
defined by $2s^\mu_N\Delta q_N=\langle N|\bar q\gamma^\mu\gamma_5
q|N\rangle$ with $s^\mu_N$ the nucleon spin four-vector.  
The parton-level Wilson coefficients $\lambda_{q}$, $\lambda_{G}$, and $\eta_{q}$
are obtained by integrating out the heavy superpartners at the matching
scale $\mu_R = m_{\widetilde{\chi}^0_1}$ and subsequently evolving down to the hadronic scale to incorporate QCD effects.
We stress that $\sigma^{\rm SD}_N$ of
Eq.~(\ref{eq:sigma-def}) is confronted directly with the LZ limits, whereas the
nuclear spin expectation values $\langle S_{p,n}\rangle$ and the associated structure
factors~\cite{Klos:2013rwa,Hu:2021awl, Hoferichter:2020osn} enter only when converting nucleon cross sections to detector 
event rates, and they are held fixed and taken to be identical
in the LZ analyses~\cite{LZ:2022lsv,LZ:2024zvo} and in the recasting
performed by \texttt{micrOMEGAs}~\cite{Belanger:2020gnr}.

Theoretical uncertainties in $\sigma^{\rm SI,SD}_N$ originate
predominantly from the coefficients $\lambda_q$, $\lambda_G$, and
$\eta_q$ ($\delta^{\rm SI,SD}_{\rm EW}$) and from the nucleon matrix
elements ($\delta^{\rm SI,SD}_{\rm Had}$), which we combine in
quadrature, $\delta^{\rm SI,SD}_\sigma=[(\delta^{\rm SI,SD}_{\rm
EW})^2+(\delta^{\rm SI,SD}_{\rm Had})^2]^{1/2}$. To reduce  $\delta_{\text{EW}}^{\text{SI,SD}}$,
we evaluate the complete one-loop corrections to the
$\widetilde\chi^0_1\widetilde\chi^0_1 h$,
$\widetilde\chi^0_1\widetilde\chi^0_1 H$ and
$\widetilde\chi^0_1\widetilde\chi^0_1 Z$ vertices (see Supplemental Material).
Box-type contributions require
either additional gaugino--Higgsino mixing insertions or heavy-sfermion
propagators; they are numerically irrelevant for a Bino-like
$\widetilde\chi^0_1$ and are therefore omitted. Owing to the strong
tree-level suppression of $C_h$ and $C_Z$, the vertex corrections shift
$\sigma^{\rm SI}_N$ by up to $50\%$ and $\sigma^{\rm SD}_N$ by up to
$30\%$, so that going beyond LO is indispensable for a reliable
prediction. We estimate the residual higher-order EW effects
by varying the matching scale over
$m_{\widetilde\chi^0_1}/2\le\mu_R\le 2m_{\widetilde\chi^0_1}$. It is found that the NLO
correction itself changes by at most $5\%$, which translates
into a shift of the cross sections of at most a few percent. The total
uncertainty is thus dominated by $\delta^{\rm SI,SD}_{\rm Had}$.
Further details are collected in the Supplemental Material.

The hadronic inputs entering  Eq.~(\ref{eq:sigma-def}) can be taken either
from lattice QCD (LQCD)~\cite{FlavourLatticeAveragingGroupFLAG:2024oxs, FlavourLatticeAveragingGroupFLAG:2021npn} or from analyses of
experimental data~\cite{Hoferichter:2015dsa,RuizdeElvira:2017stg}. We
adopt the FLAG-2024 LQCD averages~\cite{FlavourLatticeAveragingGroupFLAG:2024oxs}, namely the pion--nucleon
sigma term $\sigma_{\pi N}=(42.2\pm2.4)$~MeV, the strangeness sigma
term $\sigma_s=(44.9\pm6.4)$~MeV, and the spin fractions
$\Delta u_p=\Delta d_n=0.777\pm0.039$, $\Delta d_p=\Delta u_n=-0.438\pm0.035$ and
$\Delta s_p=\Delta s_n=-0.053\pm0.008$. The quantities $f^N_q$ and their uncertainties, which depend on
$\sigma_{\pi N}$ and $\sigma_s$, are evaluated following
Refs.~\cite{Crivellin:2013ipa,Bishara:2017pfq}. Compared with the
FLAG-2021 LQCD average~\cite{FlavourLatticeAveragingGroupFLAG:2021npn}, the precision of $\sigma_{\pi N}$
has improved by about $35\%$, from $\pm3.6$ to $\pm2.4$~MeV, while those of
$\sigma_s$ and $\Delta q_N$ are essentially unchanged. Since
$\sigma_{\pi N}$ dominates the uncertainty of $f_N$, this progress
directly sharpens the SI prediction. We note that Roy--Steiner analyses
of $\pi N$ scattering data favor a substantially larger value of $\sigma_{\pi N}$,
$\sigma_{\pi N}\simeq58$--$59$~MeV~\cite{Hoferichter:2015dsa,
Hoferichter:2015hva}, in significant tension with the LQCD
determination. In the funnel regions studied here we have verified that
a larger $\sigma_{\pi N}$ increases $\sigma^{\rm SI}_N$. The lattice value 
adopted here therefore provides a conservative choice.

Since $\sigma^{\rm SI}_N$ depends on $\sigma_{\pi N}$ and $\sigma_s$
alone, whereas $\sigma^{\rm SD}_N$ depends solely on $\Delta q_N$, the
two sets of inputs are  statistically independent and can be varied
separately. To quantify $\delta^{\rm SI,SD}_{\rm Had}$, we first
evaluate the cross sections with the central values of the inputs at
each MSSM point obtained from the scan described below, and then vary
the inputs within their $3\sigma$ ranges to determine the maximal
deviations from the central values. Collecting these deviations over
the full parameter space of Table~\ref{tab:parameters} yields the envelopes
$(-22\%,+25\%)$ for $\sigma^{\rm SI}_N$ and $(-38\%,+47\%)$ for
$\sigma^{\rm SD}_N$. The wider SD envelope reflects the partial
cancellation between the $\Delta u$ and $\Delta d$ contributions to
$g_N$, rather than a poorer precision of the inputs.

When assessing the exclusion power of LZ, we do not treat the hadronic inputs as
nuisance parameters to be marginalized as done, e.g., in
\texttt{DarkBit}~\cite{GAMBITDarkMatterWorkgroup:2017fax, GAMBIT:2018eea}. Instead, we rescale the
central-value prediction at {\bf{every}} parameter point by the most
unfavorable downward factors of the envelopes quoted above, i.e.\ $0.78$
for $\sigma^{\rm SI}_N$ and $0.62$ for $\sigma^{\rm SD}_N$. Because the
actual downward deviation at any individual point is smaller than these
global envelope factors, the cross sections entering the likelihood are
systematically below the model predictions, and the resulting exclusion
is a conservative lower bound on the true one. We emphasize that the
$3\sigma$ range invoked here refers exclusively to the variation of the
hadronic inputs; the quoted statistical significance derives from the
LZ likelihood alone, evaluated after these worst-case rescalings have
been applied.

\textit{Numerical study.---} We now examine whether the parameter
points in the funnel regions survive the LZ constraints. The scan
is guided by an objective function
$\mathcal{L}_{O}\equiv\mathcal{L}_{\Omega h^{2}}\times\mathcal{L}_{\rm DD}
\times\mathcal{L}_{\rm Const}$, where $\mathcal{L}_{\Omega h^{2}}$ is a
Gaussian centered at $(\Omega h^{2})_{c}$ with a combined theoretical
and experimental uncertainty of $\delta_{\Omega h^2}$. The direct-detection sampling
function is taken from Ref.~\cite{Matsumoto:2016hbs}:
\begin{equation}
\mathcal{L}_{\rm DD}=\exp\left[-\frac{1}{2}
\left(\frac{\sigma^{\rm SI}_{\rm eff}}{\delta\sigma^{\rm SI}}\right)^{2}
-\frac{1}{2}\left(\frac{\sigma^{\rm SD}_{\widetilde\chi^{0}_{1}-n}}
{\delta\sigma^{\rm SD}}\right)^{2}\right],
\label{eq:gaussian}
\end{equation}
where the SI term uses the effective cross section that incorporates
the natural isotopic composition of xenon, allowing a direct comparison
with the LZ results~\cite{XENON:2018voc}:
\begin{equation}
\sigma^{\rm SI}_{\rm eff}=0.169\,\sigma^{\rm SI}_{p}
+0.347\,\sigma^{\rm SI}_{n}
+0.484\sqrt{\sigma^{\rm SI}_{p}\sigma^{\rm SI}_{n}} .
\end{equation}
Its uncertainty is
$(\delta\sigma^{\rm SI})^{2}\equiv({\rm UL}^{\rm SI}_{\sigma})^{2}/1.64^{2}
+(\delta^{\rm SI}_{\sigma}\sigma^{\rm SI}_{\rm eff})^{2}$, with
${\rm UL}^{\rm SI}_{\sigma}$ the \textbf{LZ-2024} $90\%$~C.L. upper limit rescaled
by $0.12/(\Omega h^{2})_{c}$ and $\delta^{\rm SI}_{\sigma}$
parameterizing the nuclear-physics and theoretical uncertainties. The SD
term for DM--neutron scattering is constructed analogously. The penalty
function $\mathcal{L}_{\rm Const}$ equals unity when all experimental
constraints are satisfied and $e^{-100}$ otherwise. 
These constraints require: 
(i) the properties of the SM-like Higgs boson agree with the LHC
measurements at $95\%$~C.L., as evaluated with {\tt HiggsTools-1.2}
\cite{Bahl:2022igd} using $129$ Higgs observables;
(ii) the direct searches for additional Higgs bosons at LEP, the
Tevatron and the LHC, also implemented in {\tt HiggsTools-1.2}, are
satisfied;
(iii) the invisible $Z$-boson width obeys
$\Gamma_{Z\to\rm inv} \leq 2$~MeV at $95\%$~C.L. for
$m_{\widetilde\chi^{0}_{1}}\lesssim m_{Z}/2$~\cite{ALEPH:2005ab};
(iv) the $B$-physics observables ${\rm Br}(B_{s}\to\mu^{+}\mu^{-})$ and
${\rm Br}(B\to X_{s}\gamma)$ agree with the measurements within
$2\sigma$~\cite{ParticleDataGroup:2018ovx};
(v) no charge- or color-breaking minimum is present, so that only Higgs fields acquire vacuum expectation values
\cite{Alvarez-Gaume:1983drc,Gunion:1987qv,Casas:1995pd,Strumia:1996pr,
Abel:1998cc,Abel:1998wr,Cerdeno:2003yt}.
In the Higgs-data fit we assign a combined theoretical and experimental
uncertainty of $3$~GeV to $m_{h}$ and require
$\Delta\chi^{2}\equiv\chi^{2}_{\rm Higgs}-\chi^{2}_{\rm SM,125}
\lesssim6.18$, where $\chi^{2}_{\rm SM,125}\simeq154.0$ is the value we
obtain for a $125$~GeV SM Higgs boson.

\begin{table}[tbp]
	\centering
	\begin{tabular}{|c|c|c|c|c|c|}
		\hline
		\hline
		Parameter & Prior & Range & Parameter & Prior & Range \\
		\hline
		$M_1/\text{GeV}$ & Flat & $\pm\left(58\text{--}68\right)$ & $M_{\widetilde{Q}_3}/\text{TeV}$ & Flat & $1.0$--$2.5$ \\
		$\mu/\text{TeV}$ & Flat & $0.3$--$1.0$ & $M_{\widetilde{U}_3=\widetilde{D}_3}/\text{TeV}$ & Flat & $1.0$--$2.5$\\
		$M_2/\text{TeV}$ & Flat & $0.5$--$1.5$ & $A_t=A_b/\text{TeV}$ & Flat & $-6.0\text{--}6.0 $  \\
		$M_A/\text{TeV}$ & Flat & $0.8$--$2.0$ & $\tan\beta$ & Flat & $4$--$60$  \\
		\hline
		\hline
	\end{tabular}
	\caption{Scan ranges of the input parameters for the $h$-resonant
scenario, all with flat priors and defined at the renormalization scale
$Q=1$~TeV. The soft masses of the first- and second-generation squarks
and of all sleptons are fixed at $3$~TeV. The $Z$-resonant scenario is
scanned in the same way, the only difference being
$40~{\rm GeV}\le|M_{1}|\le50$~GeV. }
	\label{tab:parameters}
\end{table}

\begin{figure}[]
	\centering
	\includegraphics[scale=0.25]{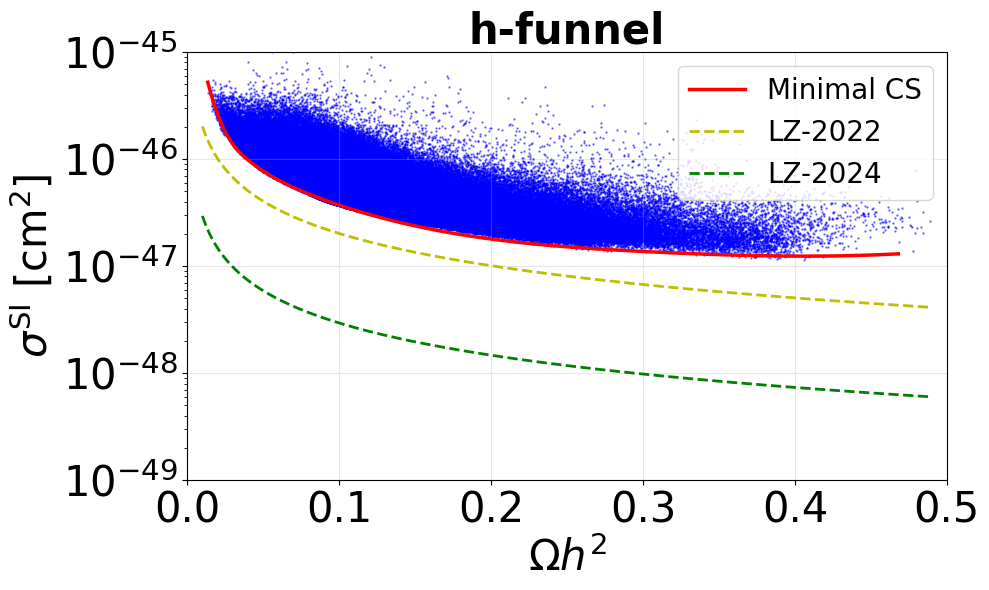} \\
    \includegraphics[scale=0.25]{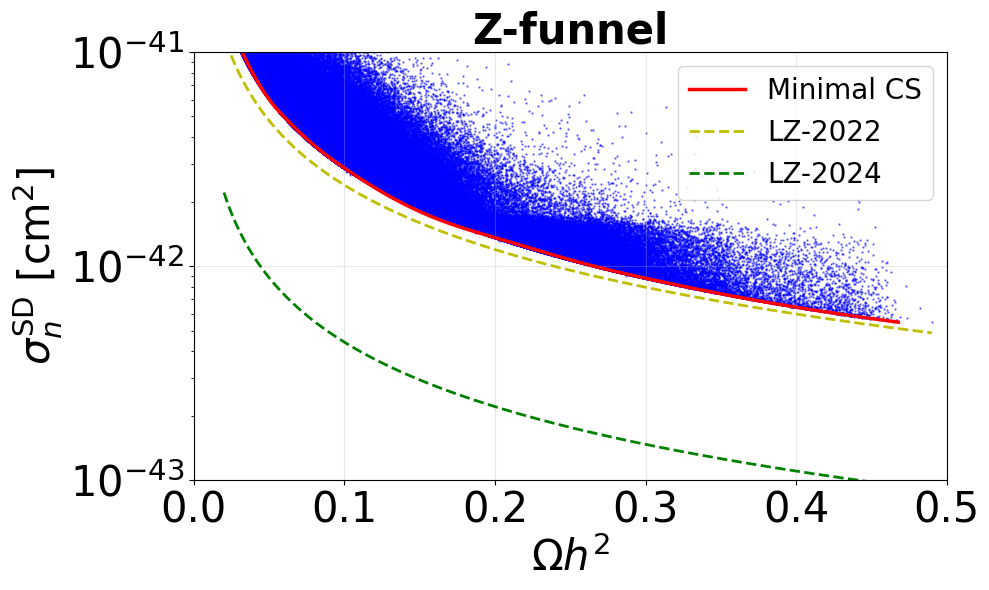}
	\caption{Variation of $\sigma^{\rm SI}$ with $\Omega h^{2}$ in the
$h$-resonance region (upper) and of $\sigma^{\rm SD}_{n}$ with
$\Omega h^{2}$ in the $Z$-resonance region (lower) for the samplings delineated in 
the text and Supplemental Material. The red curves denote the envelopes for the 
minimal cross-sections defined in the Introduction. Also shown, for comparison,
are the $90\%$~C.L. upper limits from \textbf{LZ-2022} (olive) and \textbf{LZ-2024}
(green), rescaled by $0.12/(\Omega h^{2})$. All points satisfy the
constraints specified in the text; the cross sections include the
radiative corrections and are evaluated
with the central hadronic inputs. }
	\label{fig:hzfunnel}
\end{figure}

Guided by $\mathcal{L}_{O}$, we explore the parameter space of
Table~\ref{tab:parameters} via parallelized MultiNest algorithm encoded 
in \textsf{EasyScan}~\cite{Feroz:2008xx,Shang:2023gfy} for fixed 
$\delta_{\Omega h^2} = 10\%$, $\delta^{\rm SI}_{\sigma}= \delta^{\rm SD}_{\sigma} = 20\%$,
and two representative values of $(\Omega h^2)_c = 0.06, 0.12$. 
We stress that $\mathcal{L}_{\rm DD}$ employed here is solely for efficient 
sampling of the phenomenologically viable space, not for statistical inference discussed below. 
Further details of the sampling are provided in the Supplemental Material.

For each sampled point the LZ exclusion significance is evaluated with
\texttt{micrOMEGAs}~\cite{Belanger:2026asz} for \textbf{LZ-2024} and with \texttt{DDCalc}~\cite{GAMBITDarkMatterWorkgroup:2017fax, GAMBIT:2018eea} for \textbf{LZ-2022}. The two codes differ
both in the level of detector detail retained and in the statistical
construction used to define exclusion. \texttt{micrOMEGAs} adopts a simplified
recast based on a background-free, null-observation Poisson model with
likelihood $\mathcal{L}_{\rm M}=e^{-N_{\rm ev}}$, calibrated to the
official $90\%$~C.L. limit. The expected signal count $N_{\rm ev}$ is
computed by convolving the nuclear-recoil spectrum with the detector
efficiency and energy response reported by LZ. This procedure defines the
reference $90\%$~C.L. cross sections
$\sigma^{\rm SI}_{0}(m_{\widetilde\chi^{0}_{1}})$ and
$\sigma^{\rm SD}_{0}(m_{\widetilde\chi^{0}_{1}})$~\cite{Belanger:2026asz}, such that
$N_{\rm ev}=2.3\,[\sigma^{\rm SI}/\sigma^{\rm SI}_{0}
+\sigma^{\rm SD}/\sigma^{\rm SD}_{0}] \times \Omega h^2/0.12$ at a fixed DM 
mass, with the factor $\Omega h^{2}/0.12$ accounting for the reduced local density when
$\Omega h^{2}<0.12$. Within this
construction the $p$-value is exact, $p=e^{-N_{\rm ev}}$, and is mapped
onto the one-sided Gaussian-equivalent significance $Z$ through
$p=1-\Phi(Z)$, i.e. $Z=\sqrt{2}\,{\rm Erfc}^{-1}(2p)$, with
${\rm Erfc}^{-1}$ the inverse complementary error function. 
The values $Z=1,2,3,4$ correspond to $p$-values of $15.9$, $2.28$,
$0.14$, and $0.003$, in units of $10^{-2}$, and 
$N_{\rm ev}=1.84,\,3.78,\,6.61,$ and $10.36$, respectively.

\texttt{DDCalc}, by contrast, constructs a binned Poisson likelihood from the
published \textbf{LZ-2022} efficiency and background model, profiling over a
Gaussian-constrained background normalization while fixing the
detector-response and astrophysical inputs to their central values~\cite{GAMBITDarkMatterWorkgroup:2017fax, GAMBIT:2018eea}.
With the likelihood detailed in the Supplemental Material, \texttt{DDCalc}
evaluates the test statistic
$q=-2\ln[\mathcal{L}_{\rm DDC}({\rm signal+bkg})/
\mathcal{L}_{\rm DDC}({\rm bkg\ only})]$, which by Wilks' theorem is
asymptotically distributed as $\chi^{2}_{1}$ under the background-only
hypothesis. Exclusion is calibrated to the two-sided
profile-likelihood-ratio construction employed by the LZ Collaboration,
in accordance with the reporting recommendations of Ref.~\cite{Baxter:2021pqo}.
For a uniform presentation, we convert the resulting two-sided
$p$-value into the one-sided significance used throughout this work,
$Z=\sqrt{2}\,{\rm Erfc}^{-1}[\,2\,{\rm Erfc}(\sqrt{q/2})\,]$, where 
$Z=1,2,3,4$ correspond to $q=1.99$, $5.19$, $10.27$, and $17.31$,
respectively.

\begin{figure}[]
	\centering
	\includegraphics[scale=0.20]{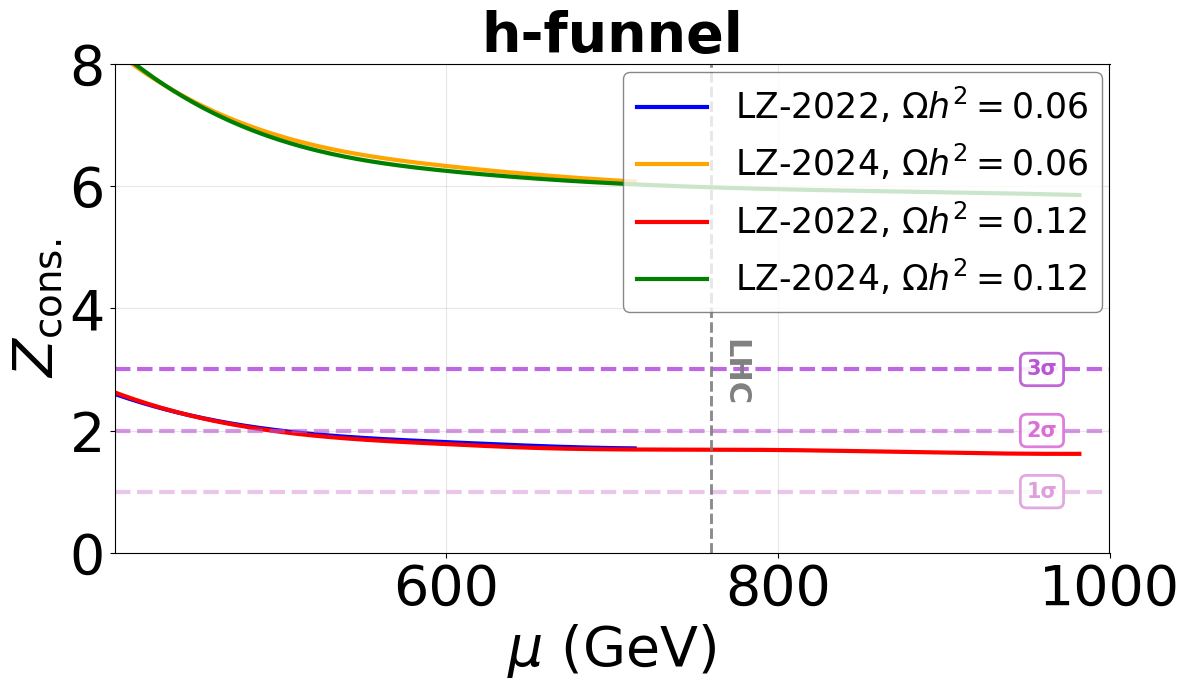} \\
    \includegraphics[scale=0.15]{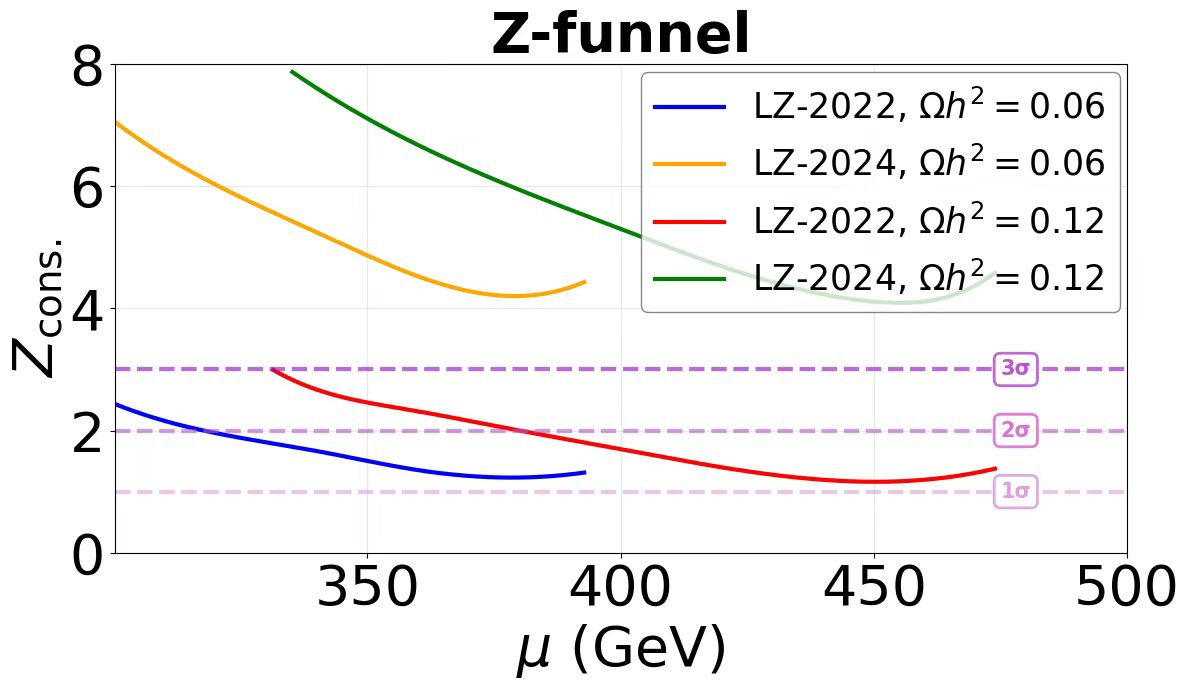}
	\caption{Minimal one-sided Gaussian-equivalent significance $Z$ versus
$\mu$ in the $h$ funnel (upper) and $Z$ funnel (lower), after applying
the adverse hadronic shifts of $-22\%$ to $\sigma^{\rm SI}$ and
$-38\%$ to $\sigma^{\rm SD}$. The \textbf{LZ-2022} and \textbf{LZ-2024} results are
evaluated with \texttt{DDCalc} and \texttt{micrOMEGAs}, respectively, for
$(\Omega h^{2})_{c}=0.06$ and $0.12$ within a $5\%$ window. The LHC EWino bound
$\mu \gtrsim 760$~GeV for the funnel regions~\cite{Athron:2026epb,Li:2025qkg}, not imposed in the analysis, is
indicated by the vertical line in the upper panel. }
	\label{fig:chi2minhzfunnel}
\end{figure}

Two caveats should be kept in mind. \texttt{DDCalc} reproduces the actual LZ
analysis as closely as the public information allows, but cannot
implement the full set of correlated nuisance parameters used internally
by the collaboration. Its significance is therefore usually
slightly optimistic. Moreover, insufficient public information for the
2024 dataset means that only the \textbf{LZ-2022} analysis is currently
implemented in \texttt{DDCalc}. The \texttt{micrOMEGAs} recast, on the other hand, does not
rely on experimental details and is calibrated only near the
$90\%$~C.L. limit, so extrapolating its idealized zero-background
Poisson tail to $3\sigma$ and beyond, although well defined, must be
regarded as a conservative estimate rather than an official experimental
significance. We have verified this point by recasting the \textbf{LZ-2022} result with
the \texttt{micrOMEGAs} method, which yields moderately smaller values of $Z$ than
\texttt{DDCalc} (see the Supplemental Material). The same tendency is apparent in
the \texttt{micrOMEGAs} recast of \textbf{LZ-2024} in Ref.~\cite{Belanger:2026asz}, where the
reference $\sigma^{\rm SI}_{0}$ exceeds the experimental limit in the
funnel region. 

Fig.~\ref{fig:hzfunnel} displays the correlation between $\Omega h^2$ and
$\sigma^{\rm SI}$ ($\sigma_n^{\rm SD}$) in the $h$-funnel
($Z$-funnel), where the red curves denote the lower envelopes of the
cross sections. These envelopes decrease with increasing
$\Omega h^{2}$, as a larger $\mu$ suppresses $C_h$ and $C_Z$, 
thereby increasing $\Omega h^2$ while reducing $\sigma^{\rm SI,SD}$.  
They lie marginally above the $90\%$~C.L. \textbf{LZ-2022} limits (olive), but exceed the
\textbf{LZ-2024} limits (green) by nearly an order of magnitude, 
placing both resonant regions in severe tension with LZ bounds.

Fig.~\ref{fig:chi2minhzfunnel} shows the minimal one-sided Gaussian-equivalent
significance $Z$ as a function of $\mu$ for both \textbf{LZ-2022} and \textbf{LZ-2024},
evaluated at $(\Omega h^{2})_{c}=0.06$ and $0.12$ within a $5\%$ window around each central value. The upper and lower
panels correspond to the $h$- and $Z$-funnel, respectively. To obtain
these most conservative exclusion envelopes, the maximal adverse
hadronic shifts, $-22\%$ for $\sigma^{\rm SI}$ and $-38\%$ for
$\sigma^{\rm SD}$, are applied to all samples before minimizing $Z$.
The envelopes decrease with $\mu$ and terminate at finite values, 
since a large $\mu$ suppresses $C_{h}$ and $C_{Z}$, and
hence $\sigma^{\rm SI,SD}$, but these couplings cannot be arbitrarily
small to reproduce $(\Omega h^{2})_{c}$ even on resonance.

Fig.~\ref{fig:chi2minhzfunnel} further shows that the envelope of $Z$ is sensitive
to $\Omega h^{2}$ for the $Z$-funnel but not for the $h$-funnel. The
origin is threefold: $C_{h}$ depends on both $\mu$ and $\tan\beta$,
whereas $C_{Z}$ is governed by $\mu$ alone; $\Omega h^{2}$ is sensitive to 
$m_{\widetilde\chi^{0}_{1}}$ in addition to the DM couplings; and
$N_{\rm ev}$ near the envelope is primarily contributed by SI (SD) scattering in the
$h$-funnel ($Z$-funnel), so that it is controlled by $\sigma^{\rm SI}$
($\sigma^{\rm SD}$) together with $\Omega h^{2}$. For the $Z$-funnel at
fixed $\mu$, $C_{Z}$ is nearly identical for the two values of
$\Omega h^{2}$, so the difference in $N_{\rm ev}$ stems
primarily from $\Omega h^{2}$. In the $h$-funnel, by contrast, $C_{h}$
can still vary appreciably at fixed $\mu$, and this variation may
compensate the difference in $\Omega h^{2}$ to yield the same
$N_{\rm ev}$.

More importantly, Fig.~\ref{fig:chi2minhzfunnel} demonstrates that the
conservative significance $Z$ obtained with \textbf{LZ-2024}
exceeds $5.9$ in the $h$-funnel and $4.1$ in the $Z$-funnel, corresponding to p-values of $1.8 \times 10^{-9}$, and $2.1 \times 10^{-5}$, respectively. 
This indicates that both resonant scenarios have been excluded. 
By contrast, the \textbf{LZ-2022} values can fall below $2$ and are 
thus insufficient for definitive exclusion. These conclusions are reached without imposing
LHC EWino constraints, which favor $\mu\gtrsim 760$~GeV
\cite{Athron:2026epb,Li:2025qkg}. Incorporating them further solidifies the exclusion
of the $Z$-funnel, where reproducing $\Omega h^{2} \leq 0.12 $ strictly requires
$\mu\lesssim 500$~GeV. Additionally, by leveraging the samples illustrated 
in Fig.~\ref{fig:hzfunnel}, we investigate parameter configurations that yield alternative 
values of the relic density. We observe that their predictions of $Z$ closely resemble 
those presented in Fig.~\ref{fig:chi2minhzfunnel}.

\textit{Conclusions.---} We have examined whether a sub-hundred-GeV Bino-dominated
$\widetilde\chi_1^0$, which attains the observed $\Omega h^2$ only through 
$h$- or $Z$-resonant annihilation in the
MSSM with heavy sfermions, survives as a thermal DM candidate after the \textbf{LZ-2024} exposure.
Since the $\widetilde\chi_1^0\widetilde\chi_1^0 h$
($\widetilde\chi_1^0\widetilde\chi_1^0 Z$) coupling controls $\Omega h^2$ and $\sigma^{\rm SI}_N$ 
($\sigma^{\rm SD}_N$) simultaneously, reproducing an
acceptable $\Omega h^2$ imposes a finite lower envelope on
$\sigma^{\rm SI}_N$ in the $h$-funnel and on $\sigma^{\rm SD}_N$ in the
$Z$-funnel. Even after including the NLO supersymmetric vertex
corrections (up to $50\%$ and $30\%$ for $\sigma^{\rm SI}_N$ and
$\sigma^{\rm SD}_N$), adopting the FLAG-2024 nucleon matrix elements
with their most adverse $3\sigma$ shifts ($-22\%$ and $-38\%$,
respectively), and performing a likelihood-based recast of the LZ data,
we find that these envelopes lie well above the \textbf{LZ-2024} bounds. 
The conservative one-sided Gaussian-equivalent significance exceeds $5.9$ in
the $h$-funnel and $4.1$ in the $Z$-funnel, corresponding to $p$-values
of $1.8\times10^{-9}$ and $2.1\times10^{-5}$, respectively. Both resonant regions are
therefore closed, whether $\widetilde\chi_1^0$ saturates the observed
abundance or constitutes only a subdominant component, and independently
of LHC EWino searches. These findings substantially sharpen
earlier analyses based on the SI results of \textbf{LZ-2022} and LHC EWino searches~\cite{Barman:2022jdg}, which retained viable $h$-funnel points for $M_1>0$ and
blind-spot regions for $M_1<0$, in agreement with the
$Z<2$ values we obtain for \textbf{LZ-2022}. 

Although derived within the MSSM, the argument applies to any Majorana DM
candidate whose relic abundance and nucleon scattering originate from the
same $h$- or $Z$-mediated interaction, provided that the scattering
amplitude is not accidentally suppressed by cancellations against other
contributions and that no additional annihilation or entropy-injection
channels intervene. The basics is that the LZ limits require the mediator coupling to be
feeble, whereas the relic abundance forbids it from being arbitrarily
small. With the funnels closed, a thermal Bino-like $\widetilde\chi^0_1$ in the
MSSM must resort to coannihilation, heavy-Higgs resonance, or a
well-tempered Bino--Higgsino mixture. Combined with the LHC searches for
additional Higgs bosons~\cite{ATLAS:2020zms,CMS:2022goy} and for EWinos and sleptons in compressed
spectra~\cite{ATLAS:2021moa,ATLAS:2025evx}, and with the stringent direct-detection limits on the
Bino--Higgsino mixing, these options favor
$|m_{\widetilde\chi^0_1}|\gtrsim200$~GeV, apart from exceptional
configurations such as the weakly constrained $\widetilde\tau$-coannihilation
strip~\cite{Bagnaschi:2017tru,ATLAS:2026gdo}.

\begin{acknowledgments}
\textit{Acknowledgements.}---
This work resulted from extensive and in-depth discussions among all three authors, 
who are listed alphabetically by surname. J.~Cao conceived the research project and 
performed the parameter-space scans. S.~Bisal developed the numerical codes 
for the radiative corrections and quantified the uncertainties from hadronic 
matrix elements. Z.~Bao analyzed the results and prepared the figures. 
The manuscript was written jointly by S.~Bisal and J.~Cao.

We thank Pengxuan Zhu for helpful discussions. This work was supported by 
the National Natural Science Foundation of China (NSFC) under Grant No.~12575110.
\end{acknowledgments}





%

\clearpage

\appendix

\onecolumngrid

\setcounter{figure}{0} 
\setcounter{page}{1}

\begin{figure}[]
	\centering
	\newcommand{\scl}{0.60}
	\newcommand{\lnwdth}{0.7pt}
	\newcommand{\arsize}{1pt}  
	
	\begin{tikzpicture}[scale=\scl]
		\begin{feynman}[every edge/.style={line width=\lnwdth}]
			\vertex (a1) at (-0.4,1.4) {\(\widetilde{\chi}^0_1\)};
			\vertex (a2) at (-0.2,-1.4) {\(q\)};
			\vertex (a3) at (2.3,1.4) {\(\widetilde{\chi}^0_1\)};
			\vertex (a4) at (2.25,-1.4) {\(q\)};
			\vertex (b) at (1,0.4);
			\vertex (c) at (1,-0.5);
			
			\diagram*{ 
				(a1) -- [fermion, arrow size=\arsize] (b) -- [fermion, arrow size=\arsize] (a3),
				(a2) -- [fermion, arrow size=\arsize] (c) -- [fermion, arrow size=\arsize] (a4),
				(b) -- [scalar, edge label=$h_i/Z$] (c),
				(a1) -- [photon] (b),
				(b) -- [photon] (a3),
			};
			\node at (1,-2.02) {(a)};
		\end{feynman}
	\end{tikzpicture}
	\hspace{0.1cm}  
	\begin{tikzpicture}[scale=\scl]
		\begin{feynman}[every edge/.style={line width=\lnwdth}]
			\vertex (n1) at (0.5,1.4) {\(\widetilde{\chi}^0_1\)};
			\vertex (n2) at (3.5,1.4) {\(\widetilde{\chi}^0_1\)};
			\vertex (v1) at (2,0.6);
			\vertex (v2) at (2,0.2);
			\vertex (b1) at (1.3,-0.5);
			\vertex (b2) at (2.7,-0.5);
			\vertex (b3) at (2,0.2);
			\vertex (g1) at (0.6,-1.5) {\(g\)};
			\vertex (g2) at (3.4,-1.5) {\(g\)};
			
			\diagram*{
				(n1) -- [fermion, arrow size=\arsize] (v1) -- [fermion, arrow size=\arsize] (n2),
				(v1) -- [scalar, edge label=$h_i$] (v2),
				(v2) -- [fermion, arrow size=\arsize] (b3),
				(b1) -- [fermion, arrow size=\arsize] (b2)
				-- [fermion, arrow size=\arsize] (b3)
				-- [fermion, arrow size=\arsize] (b1),
				(b1) -- [gluon] (g1),
				(b2) -- [gluon] (g2),
				(n1) -- [photon] (v1),
				(v1) -- [photon] (n2),
			};
			\node at (2,-2.02) {(b)};
		\end{feynman}
	\end{tikzpicture}
	
	\caption{\justifying LO $\widetilde{\chi}_1^0$--$(q,g)$ interactions: (a) $h_i$- and $Z$-mediated $\widetilde{\chi}_1^0$--$q$ scattering; (b) Heavy quark loop inducing $\widetilde{\chi}_1^0$--$g$ scattering.}
	\label{fig:sisdschematic}
\end{figure}
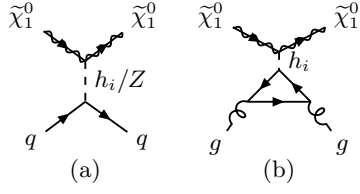

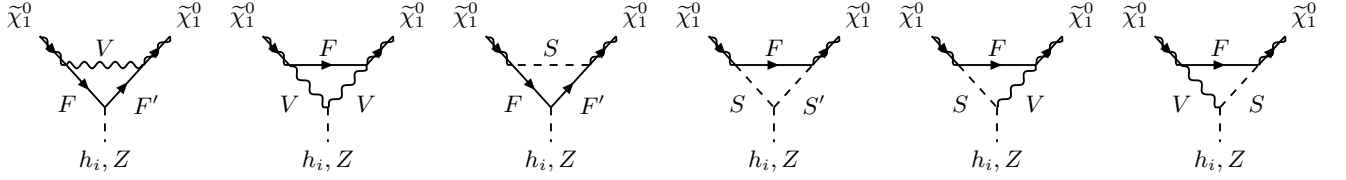
\begin{figure}[]
	\newcommand{\scl}{0.51}
	\newcommand{\Xloc}{2.5,-2.1}
	\newcommand{\labloc}{2.5,-3}
	\newcommand{\bbbc}{2.5, -0.7}
	\newcommand{\bbbb}{3.5,0.4}
	\newcommand{\bbba}{1.5, 0.4}
	\newcommand{\lnwdth}{0.7pt}
	
	\centering
	\begin{tabular}{cccccc}  
		\begin{tikzpicture}[scale=\scl]
			\begin{feynman}[every edge/.style={line width=\lnwdth}]
				\vertex (a1) at (0.3,1.7) {\(\widetilde{\chi}^0_1\)};
				\vertex (b1) at (\bbba);
				\vertex (b2) at (\bbbb);
				\vertex (b3) at (\bbbc);
				\vertex (a3) at (4.7,1.7) {\(\widetilde{\chi}^0_1\)};
				\vertex (c) at (\Xloc) {$h_i,Z$};
				
				\diagram*{
					(a1) -- [fermion, arrow size=1pt] (b1),
					(a1) -- [photon] (b1),   
					(b1) -- [fermion, edge label'=$F$, arrow size=1pt] (b3)
					-- [fermion, edge label'=$F'$, arrow size=1pt] (b2),
					(b2) -- [fermion, arrow size=1pt] (a3),
					(b2) -- [photon] (a3),   
					(b2) -- [photon, edge label'=$V$] (b1),
					(b3) -- [scalar] (c),
				};
				\node at (\labloc) {};
			\end{feynman}
		\end{tikzpicture}
		&
		\begin{tikzpicture}[scale=\scl]
			\begin{feynman}[every edge/.style={line width=\lnwdth}]
				\vertex (a1) at (0.3,1.7) {\(\widetilde{\chi}^0_1\)};
				\vertex (b1) at (\bbba);
				\vertex (b2) at (\bbbb);
				\vertex (b3) at (\bbbc);
				\vertex (a3) at (4.7,1.7) {\(\widetilde{\chi}^0_1\)};
				\vertex (c) at (\Xloc) {$h_i,Z$};
				
				\diagram*{
					(a1) -- [fermion, arrow size=1pt] (b1),
					(a1) -- [photon] (b1),
					(b1) -- [photon, edge label'=$V$] (b3)
					-- [photon, edge label'=$V$] (b2),
					(b2) -- [fermion, arrow size=1pt] (a3),
					(b2) -- [photon] (a3),
					(b1) -- [fermion, edge label=$F$, arrow size=1pt] (b2),
					(b3) -- [scalar] (c),
				};
				\node at (\labloc) {};
			\end{feynman}
		\end{tikzpicture}
		&
		\begin{tikzpicture}[scale=\scl]
			\begin{feynman}[every edge/.style={line width=\lnwdth}]
				\vertex (a1) at (0.3,1.7) {\(\widetilde{\chi}^0_1\)};
				\vertex (b1) at (\bbba);
				\vertex (b2) at (\bbbb);
				\vertex (b3) at (\bbbc);
				\vertex (a3) at (4.7,1.7) {\(\widetilde{\chi}^0_1\)};
				\vertex (c) at (\Xloc) {$h_i,Z$};
				
				\diagram*{
					(a1) -- [fermion, arrow size=1pt] (b1),
					(a1) -- [photon] (b1),
					(b1) -- [fermion, edge label'=$F$, arrow size=1pt] (b3)
					-- [fermion, edge label'=$F'$, arrow size=1pt] (b2),
					(b2) -- [fermion, arrow size=1pt] (a3),
					(b2) -- [photon] (a3),
					(b2) -- [scalar, edge label'=$S$] (b1),
					(b3) -- [scalar] (c),
				};
				\node at (\labloc) {};
			\end{feynman}
		\end{tikzpicture}
		&
		\begin{tikzpicture}[scale=\scl]
			\begin{feynman}[every edge/.style={line width=\lnwdth}]
				\vertex (a1) at (0.3,1.7) {\(\widetilde{\chi}^0_1\)};
				\vertex (b1) at (\bbba);
				\vertex (b2) at (\bbbb);
				\vertex (b3) at (\bbbc);
				\vertex (a3) at (4.7,1.7) {\(\widetilde{\chi}^0_1\)};
				\vertex (c) at (\Xloc) {$h_i,Z$};
				
				\diagram*{
					(a1) -- [fermion, arrow size=1pt] (b1),
					(a1) -- [photon] (b1),
					(b1) -- [scalar, edge label'=$S$] (b3)
					-- [scalar, edge label'=$S'$] (b2),
					(b2) -- [fermion, arrow size=1pt] (a3),
					(b2) -- [photon] (a3),
					(b1) -- [fermion, edge label=$F$, arrow size=1pt] (b2),
					(b3) -- [scalar] (c),
				};
				\node at (\labloc) {};
			\end{feynman}
		\end{tikzpicture}
		&
		\begin{tikzpicture}[scale=\scl]
			\begin{feynman}[every edge/.style={line width=\lnwdth}]
				\vertex (a1) at (0.3,1.7) {\(\widetilde{\chi}^0_1\)};
				\vertex (b1) at (\bbba);
				\vertex (b2) at (\bbbb);
				\vertex (b3) at (\bbbc);
				\vertex (a3) at (4.7,1.7) {\(\widetilde{\chi}^0_1\)};
				\vertex (c) at (\Xloc) {$h_i,Z$};
				
				\diagram*{
					(a1) -- [fermion, arrow size=1pt] (b1),
					(a1) -- [photon] (b1),
					(b1) -- [scalar, edge label'=$S$] (b3)
					-- [photon, edge label'=$V$] (b2),
					(b2) -- [fermion, arrow size=1pt] (a3),
					(b2) -- [photon] (a3),
					(b1) -- [fermion, edge label=$F$, arrow size=1pt] (b2),
					(b3) -- [scalar] (c),
				};
				\node at (\labloc) {};
			\end{feynman}
		\end{tikzpicture}
		&
		\begin{tikzpicture}[scale=\scl]
			\begin{feynman}[every edge/.style={line width=\lnwdth}]
				\vertex (a1) at (0.3,1.7) {\(\widetilde{\chi}^0_1\)};
				\vertex (b1) at (\bbba);
				\vertex (b2) at (\bbbb);
				\vertex (b3) at (\bbbc);
				\vertex (a3) at (4.7,1.7) {\(\widetilde{\chi}^0_1\)};
				\vertex (c) at (\Xloc) {$h_i,Z$};
				
				\diagram*{
					(a1) -- [fermion, arrow size=1pt] (b1),
					(a1) -- [photon] (b1),
					(b1) -- [photon, edge label'=$V$] (b3)
					-- [scalar, edge label'=$S$] (b2),
					(b2) -- [fermion, arrow size=1pt] (a3),
					(b2) -- [photon] (a3),
					(b1) -- [fermion, edge label=$F$, arrow size=1pt] (b2),
					(b3) -- [scalar] (c),
				};
				\node at (\labloc) {};
			\end{feynman}
		\end{tikzpicture}
	\end{tabular}
	
	\caption{\justifying  Diagrams for one-loop virtual correction to the $\widetilde{\chi}_1^0\widetilde{\chi}_1^0h_i$ and $\widetilde{\chi}_1^0\widetilde{\chi}_1^0Z$ vertices. The internal particles are MSSM states: $V\in\{W^\pm,Z\}$, $F(F')\in\{\widetilde{\chi}^0_{1\dots4},\widetilde{\chi}_{1,2}^\pm,\ell,\nu_\ell,q\}$ , and $S(S')\in\{h,H,A,H^\pm,G^{0,\pm},\widetilde\ell,\widetilde\nu_{\ell},\widetilde q\}$ . Box diagrams also contribute to the DM--nucleon scattering but their effects are usually negligible in the heavy sfermion limit for nearly pure Bino DM.}
	\label{fig:topologies}
\end{figure}

\section*{Supplemental Material}

\subsection*{Computation of the EW corrections: DM--nucleon scattering at NLO}

\paragraph{1. LO couplings.}
The interactions of the LSP $\widetilde\chi^0_1$ with the CP-even Higgs bosons $h_i\in\{h,H\}$ and the $Z$ boson are
\begin{equation}
\mathcal L_{\widetilde\chi^0_1\widetilde\chi^0_1 h_i}\supset
-\tfrac12\, h_i\,\overline{\widetilde\chi^0_1}\big[C^{h_i}_L P_L+C^{h_i}_R P_R\big]\widetilde\chi^0_1 ,
\qquad
\mathcal L_{\widetilde\chi^0_1\widetilde\chi^0_1 Z}\supset
\tfrac12\, Z_\mu\,\overline{\widetilde\chi^0_1}\gamma^\mu\big[C^{Z}_L P_L+C^{Z}_R P_R\big]\widetilde\chi^0_1 ,
\label{eq:Lint}
\end{equation}
with $P_{L,R}=(1\mp\gamma_5)/2$. For a Bino-like $\widetilde\chi^0_1$ in the limit $|M_1|\ll|M_2|,|\mu|$, the LO couplings reduce to
\begin{equation}
C^{\rm LO}_{h}\simeq-\frac{g_2 s_W t_W m_Z}{\mu^2-M_1^2}\,(\mu s_{2\beta}+M_1),\qquad
C^{\rm LO}_{H}\simeq\frac{g_2 s_W t_W m_Z}{\mu^2-M_1^2}\,\mu c_{2\beta},\qquad
C^{\rm LO}_{Z}\simeq\pm\frac{g_2}{2c_W}\,\frac{s_W^2 m_Z^2}{\mu^2-M_1^2}\,c_{2\beta},
\label{eq:LOcouplings}
\end{equation}
where $C^{h_i,{\rm LO}}_L=C^{h_i,{\rm LO}}_R\equiv C^{\rm LO}_{h_i}$, $|C^{Z,{\rm LO}}_L|=|C^{Z,{\rm LO}}_R|\equiv|C^{\rm LO}_Z|$, and the upper (lower) sign of $C^{\rm LO}_Z$ applies to the left- (right-)handed component. Here $\tan\beta=v_u/v_d$ is the ratio of the vacuum expectation values of the up- and down-type Higgs doublets, $s_W$, $c_W$, and $t_W$ denote the sine, cosine, and tangent of the weak mixing angle $\theta_W$, and $c_x\equiv\cos x$, $s_x\equiv\sin x$. These
expressions exhibit the suppression of the tree-level couplings
as the Higgsino admixture decreases. The $Z$ coupling is
axial-vector, as required for a diagonal Majorana-neutralino
interaction. Note that $C^{\rm LO}_h$ vanishes at $M_1=-\mu s_{2\beta}=-2\mu t_\beta/(1+t_\beta^2)$, the LO blind spot of SI direct detection.

\vspace{0.2cm}

\paragraph{2. Wilson coefficients.} Figs.~\ref{fig:sisdschematic}(a) and \ref{fig:sisdschematic}(b)
show the LO diagrams for $\widetilde\chi^0_1$--$q$ and $\widetilde\chi^0_1$--$g$ scattering in the heavy-sfermion limit, 
from which the coefficients in Eq.~\eqref{eq:L_eff} follow as~\cite{Drees:1993bu}
\begin{equation}
\lambda_q=\sum_{h_i=h,H}\frac{C_{h_i}C^q_{h_i}}{2m_{h_i}^2},\qquad
\lambda_G=-\frac{\alpha_s}{12\pi}\sum_{Q=c,b,t}c_Q\lambda_Q,\qquad
\eta_q=-\frac{g_2 c_W}{4m_W^2}\,T^q_3\,C_Z ,
\label{eq:Wilson}
\end{equation}
where $C_{h_i(Z)}=C^{\rm LO}_{h_i(Z)}$ at LO, $C^q_{h_i}=g_2 y^q_{h_i}/(2m_W)$ with $y^q_{h_i}$ the quark Yukawa couplings normalized to their SM values~\cite{Djouadi:2005gj}, $c_Q=1+11\alpha_s(m_Q)/(4\pi)$ accounts for the QCD correction to the heavy-quark loop, $\lambda_Q$ is defined analogously to $\lambda_q$, and $T^q_3=+1/2\,(-1/2)$ for up- (down-) type quarks. Over most of the parameter space considered here, the $H$-mediated contribution is strongly suppressed relative to the $h$-mediated one by the larger $H$ mass. For a predominantly Bino-like $\widetilde\chi^0_1$ with $|N_{11}|^2>99\%$, both $C^{\rm LO}_h$ and $C^{\rm LO}_Z$ are of $\mathcal O(10^{-3})$ in their respective resonant scenarios.

\vspace{0.2cm}

\paragraph{3. NLO vertex corrections.}
The diagrams in Figs.~\ref{fig:sisdschematic}(a) and \ref{fig:sisdschematic}(b) receive potentially significant one-loop corrections to the $\widetilde\chi^0_1\widetilde\chi^0_1 h_i$ and $\widetilde\chi^0_1\widetilde\chi^0_1 Z$ vertices. Box diagrams also contribute to the scattering amplitude, but for a nearly pure Bino they are suppressed either by the heavy first- and second-generation sfermion masses or by the small Wino and Higgsino fractions of $\widetilde\chi^0_1$, and are therefore neglected. The virtual corrections are classified into the six generic triangle topologies shown in Fig.~\ref{fig:topologies}, with internal vector bosons $V\in\{W^\pm,Z\}$, fermions $F,F'\in\{\widetilde\chi^0_{1\ldots4},\widetilde\chi^\pm_{1,2},\ell,\nu_\ell,q\}$, and scalars $S,S'\in\{h,H,A,H^\pm,G^{0,\pm},\widetilde\ell,\widetilde\nu_\ell,\widetilde q\}$, which exhaust the MSSM contributions at one loop. In total $904$ diagrams are evaluated: $306$ for the $\widetilde\chi^0_1\widetilde\chi^0_1 h$ vertex, $306$ for $\widetilde\chi^0_1\widetilde\chi^0_1 H$, and $292$ for $\widetilde\chi^0_1\widetilde\chi^0_1 Z$. Several of these topologies are UV divergent. We renormalize the vertices 
following the on-shell scheme of Refs.~\cite{Fritzsche:2002bi,Frank:2006yh,Fritzsche:2013fta,Sperling:2013eva} for the EWino and Higgs sectors. The renormalized couplings read
\begin{equation}
C^{\rm NLO}_{h_i(Z)}=C^{\rm LO}_{h_i(Z)}+C^{\rm 1L}_{h_i(Z)}+\delta C_{h_i(Z)},
\label{eq:CNLO}
\end{equation}
where $C^{\rm 1L}_{h_i(Z)}$ and $\delta C_{h_i(Z)}$ denote the sum of the one-loop virtual diagrams and the counterterm contribution, respectively. Technical details of the renormalization procedure can be found in 
Refs.~\cite{Bisal:2023fgb, Bisal:2023iip, Bisal:2024ezn, Bisal:2024iar}.

The relative importance of the diagrams in  Fig.~\ref{fig:topologies} are as follows:
\begin{itemize}
\item For the $\widetilde\chi^0_1\widetilde\chi^0_1 h$ vertex, the first two topologies are suppressed by the small Wino and Higgsino fractions and contribute at $\mathcal O(10^{-7})$--$\mathcal O(10^{-6})$. In the third and fourth topologies, Higgs-boson loops are suppressed by the small Higgsino--Higgsino--Higgs coupling, and slepton loops by the heavy slepton masses, the absence of a color factor, and the small chiral mixing; squark loops, by contrast, dominate. Third-generation quark--squark loops, with large Yukawa couplings and moderate squark masses, yield $\mathcal O(10^{-4})$--$\mathcal O(10^{-3})$. The last two topologies reach at most $\mathcal O(10^{-5})$, again owing to the small Wino and Higgsino fractions.

\item For the $\widetilde\chi^0_1\widetilde\chi^0_1 Z$ vertex, the first two and the last two topologies, as well as the slepton loops in the third and fourth topologies, are suppressed for the same reasons. Higgs-boson loops in the third and fourth topologies are also unimportant, in this case because the relevant couplings are suppressed at large $\tan\beta$~\cite{Calibbi:2014lga}. Squark loops again dominate. Since the Bino--quark--squark couplings are flavor universal, no pronounced third-generation enhancement arises; the chiral mixing in the stop and sbottom sectors, together with their comparatively light masses, nevertheless produces a mild enhancement, giving $\mathcal O(10^{-3})$ overall.
 \end{itemize}

\vspace{0.2cm}

\paragraph{4. Implementation in \texttt{micrOMEGAs}.}
The NLO-improved cross sections are obtained via the following pipeline:
\begin{itemize}
\item The one-loop diagrams for the $\widetilde\chi^0_1\widetilde\chi^0_1 h_i$ and $\widetilde\chi^0_1\widetilde\chi^0_1 Z$ vertices are generated with \texttt{FeynArts~3.11}~\cite{Fritzsche:2013fta,Hahn:2000kx,Hahn:2001rv} in the Feynman gauge, and the amplitudes are reduced to Passarino--Veltman integrals with \texttt{FormCalc~9.9}~\cite{Fritzsche:2013fta,Hahn:1998yk}.
    
\item The counterterms are fixed in the $\mathrm{CCN[1]}$ on-shell scheme~\cite{Fritzsche:2013fta,Chatterjee:2011wc}, in which both charginos $\widetilde\chi^\pm_{1,2}$ and the lightest neutralino $\widetilde\chi^0_1$ are renormalized on shell, as implemented in \texttt{FormCalc}~\cite{Fritzsche:2013fta}.
  
\item The renormalized vertices are evaluated numerically with \texttt{LoopTools~2.16}~\cite{Hahn:1998yk}, using the MSSM spectrum generated by \texttt{SPheno~4.0.5}~\cite{Porod:2003um,Staub:2017jnp} from the model file of \texttt{SARAH~4.14.5}~\cite{Staub:2017jnp,Staub:2013tta,Staub:2015kfa}. Ultraviolet finiteness is explicitly verified for all sampled points.

\item The LO SI and SD cross sections, dominated by the diagrams in Fig.~\ref{fig:sisdschematic}, are computed with  {\tt micrOMEGAs}~6.2.4 \cite{Belanger:2001fz, Belanger:2006is, Belanger:2008sj, Belanger:2013oya, Alguero:2023zol}. The $\widetilde\chi^0_1\widetilde\chi^0_1 h_i$ and $\widetilde\chi^0_1\widetilde\chi^0_1 Z$ vertices in \texttt{micrOMEGAs} are then replaced by $C^{\rm NLO}_{h_i(Z)}$ of Eq.~\eqref{eq:CNLO}. With these vertices, \texttt{micrOMEGAs} evaluates, for each parameter point, the SI and SD cross sections and the annihilation cross sections entering the relic density [cf.\ Eq.~\eqref{eq:sigma_hZ}] at NLO. We realize that the NLO effective coupling expanded to order $\mathcal{O}(g_2^4)$ is given by
\begin{equation}
|C_{h(Z)}^{\rm NLO}|^2 = |C_{h(Z)}^{\rm LO}|^2 + 2\,\mathrm{Re}\left[ C_{h(Z)}^{\rm LO} \left( C_{h(Z)}^{\rm 1L} + \delta C_{h(Z)} \right)^* \right] + \mathcal{O}(g_2^6),
\end{equation}
our treatment therefore includes part of the NNLO effects.
\end{itemize}

\vspace{0.2cm}

\begin{figure}[]
	\centering
	\includegraphics[width=0.49\linewidth]{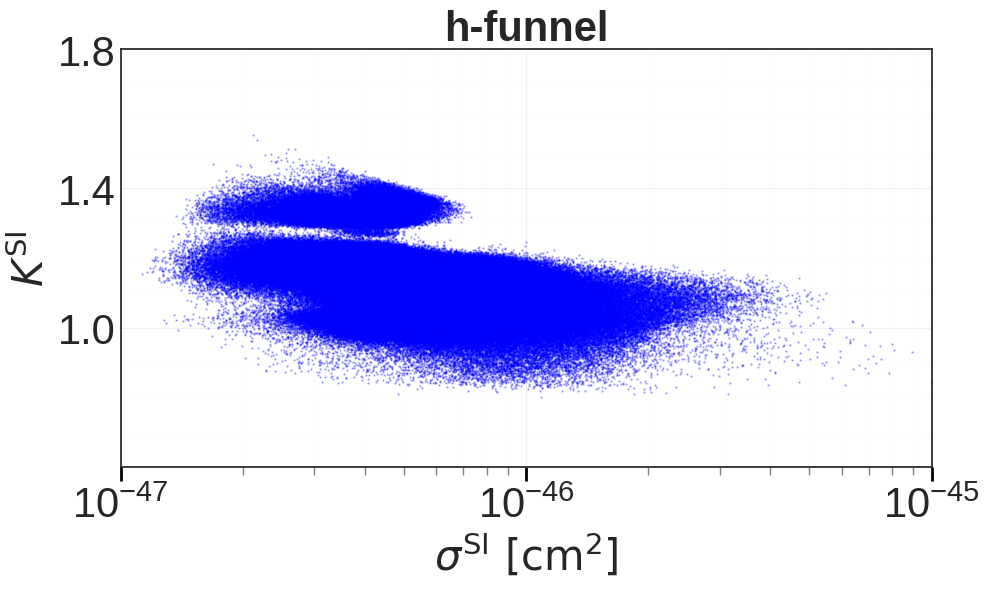}
	\includegraphics[width=0.49\linewidth]{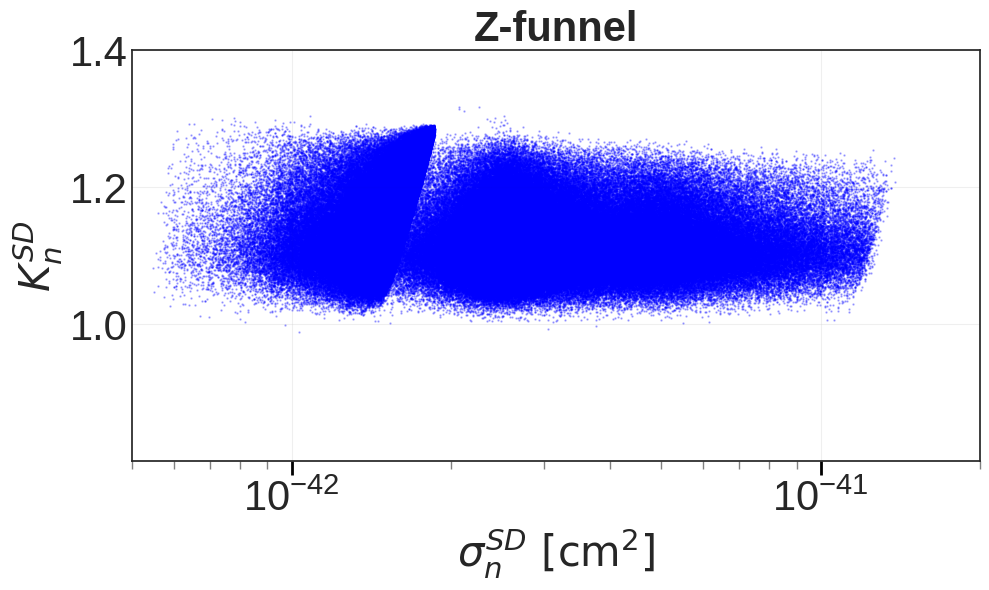}\\
	\caption{Left: Variation of $K^{\rm SI}$ with $\sigma^{\rm SI}$ in the $h$-resonant scenario. Right: Variation of $K_n^{\rm SD}$ with $\sigma_n^{\rm SD}$ in the $Z$-resonant scenario. All points reproduce the observed relic abundance at $2\sigma$ level and satisfy the Higgs data constraints, $B$-physics observables, and the absence of CCB minima.}
	\label{KSISD}
\end{figure}

\paragraph{5. Magnitude of the NLO corrections.}
We quantify the corrections to the scattering cross-sections by the $K$ factor, $K\equiv\sigma^{\rm NLO}/\sigma^{\rm LO}$, evaluated for $K^{\rm SI}$ and $K^{\rm SD}_n$ in the $h$- and $Z$-funnel scenarios, respectively, over all parameter points that satisfy the Higgs-data fit, the $B$-physics constraints, and the absence of CCB minima. As illustrated in Fig.~\ref{KSISD}, $K^{\rm SI}\lesssim1.5$ and $K^{\rm SD}_n\lesssim1.3$, translating to a maximum 
shift relative to the LO prediction by approximate $50\%$ and $30\%$, respectively. These largest corrections arise when both the lighter stop and the Higgsinos are moderately light and their contributions add constructively, and $K_n^{\rm SD}$ is further enhanced if one assumes moderately lighter first-two-generation squarks. We note that the corrections become large when the Higgsino-like and Wino-like neutralinos are nearly degenerate; this growth reflects the numerical instability of the on-shell mixing counterterms for quasi-degenerate states rather than a physical effect, and we exclude it by requiring a mass splitting of at least $3~\mathrm{GeV}$ between the Higgsino-like and Wino-like neutralinos. We stress that the shifts are different from the uncertainties remaining after the one-loop vertex corrections are included. As described in the main text, varying the scale used to estimated missing higher-order effects changes the one-loop corrections by at most about $5\%$, corresponding to a shift of at most a few percent in the cross sections. 

\subsection*{Computation of the relic density}

Assuming that $\widetilde\chi^0_1$ is a thermal relic, its annihilation cross section into SM fermion pairs is~\cite{Hamaguchi:2015rxa}
\begin{equation}
\sigma(\widetilde\chi^0_1\widetilde\chi^0_1\to f\bar f)=\sigma_h+\sigma_Z+\ldots,
\label{eq:sigma_tot}
\end{equation}
where the $h$- and $Z$-exchange contributions read
\begin{equation}
\sigma_{h(Z)}=\big|C^{\rm NLO}_{h(Z)}\big|^2\,\frac{1}{\xi}
\left(1-\frac{4m_{\widetilde\chi^0_1}^2}{s}\right)^{\frac{1}{2}}
\frac{s}{m_{h(Z)}}\,
\frac{\Gamma_{h(Z)\to f\bar f}}{\big(s-m_{h(Z)}^2\big)^2+\big(m_{h(Z)}\Gamma_{h(Z)}\big)^2}.
\label{eq:sigma_hZ}
\end{equation}
Here $\xi=2\,(1)$ for $h\,(Z)$ exchange, 
$\Gamma_{h(Z)\to f\bar f}$ and $\Gamma_{h(Z)}$ are the partial and total widths of the mediator, $\sqrt s$ is the 
center-of-mass energy, and terms of $\mathcal O(m_f^2/m_{\widetilde\chi^0_1}^2)$ are neglected. The ellipsis in 
Eq.~\eqref{eq:sigma_tot} represents $H/A$ exchange, $t$-channel sfermion exchange, box topologies, and 
interference terms, all of which are strongly suppressed in the heavy-sfermion, Bino-dominated regime 
under LHC constraints. The cross sections $\sigma_h$ and $\sigma_Z$ are resonantly enhanced at 
$s\simeq m_h^2$ and $s\simeq m_Z^2$, which for non-relativistic DM at freeze-out corresponds to 
$m_{\widetilde\chi^0_1}\simeq m_h/2$ and $m_Z/2$; in these regions $\sigma_h$ or $\sigma_Z$ completely dominates 
the total annihilation cross section.

The relic density is governed by the thermally averaged cross section~\cite{Arcadi:2017kky}
\begin{equation}
\langle\sigma v_{\rm rel}\rangle(T)=
\frac{1}{8m_{\widetilde\chi^0_1}^4\,T\,\big[K_2(m_{\widetilde\chi^0_1}/T)\big]^2}
\int_{4m_{\widetilde\chi^0_1}^2}^{\infty}\!ds\;\sigma(s)\,\big(s-4m_{\widetilde\chi^0_1}^2\big)\sqrt s\,K_1\!\left(\frac{\sqrt s}{T}\right),
\label{eq:thermal_avg}
\end{equation}
where $K_{1,2}$ are modified Bessel functions of the second kind and $\sigma(s)\simeq\sigma_{h(Z)}$ in the $h\,(Z)$ funnel. The integral over $s$ accounts for the thermal smearing of the initial-state momenta, while the finite width $\Gamma_{h(Z)}$ keeps $\sigma(s)$ finite on resonance. Together, these two effects bound $\langle\sigma v_{\rm rel}\rangle$ from above for given couplings and thus set the lower envelopes on $\sigma^{\rm SI}$ and $\sigma^{\rm SD}_n$ discussed in the main text. Solving the Boltzmann equation from freeze-out to the present yields, in the standard approximation~\cite{Jungman:1995df,Bertone:2004pz},
\begin{equation}
\Omega h^2\simeq0.12\left(\frac{80}{g_*}\right)^{1/2}\left(\frac{x_F}{25}\right)
\left(\frac{2.3\times10^{-26}\,\mathrm{cm^3\,s^{-1}}}{\langle\sigma v_{\rm rel}\rangle_{x_F}}\right),
\label{eq:omega}
\end{equation}
where $x_F=m_{\widetilde\chi^0_1}/T_F$ with $T_F$ the freeze-out temperature and $g_*$ the effective number of relativistic degrees of freedom at $T_F$. Equation~\eqref{eq:omega} serves only to illustrate the scaling of $\Omega h^2$ with $\langle\sigma v_{\rm rel}\rangle$; since $\langle\sigma v_{\rm rel}\rangle$ varies rapidly with $T$ near a resonance, all numerical results in this work are obtained by solving the Boltzmann equation numerically with \texttt{micrOMEGAs}, using the NLO vertices of Eq.~\eqref{eq:CNLO} and including all kinematically accessible annihilation processes.

\begin{figure}[]
	\centering
	\includegraphics[width=0.44\linewidth]{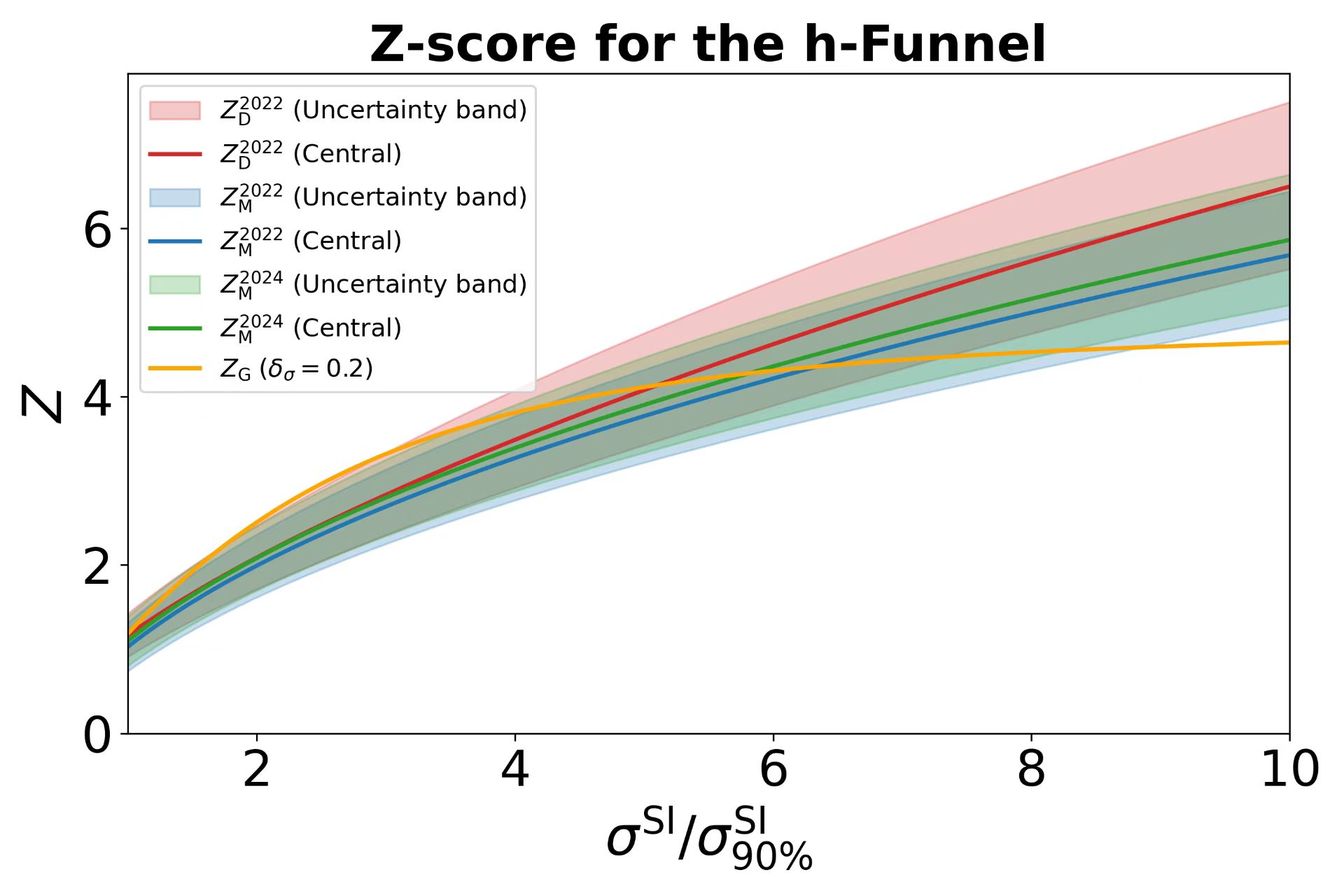}
	\includegraphics[width=0.44\linewidth]{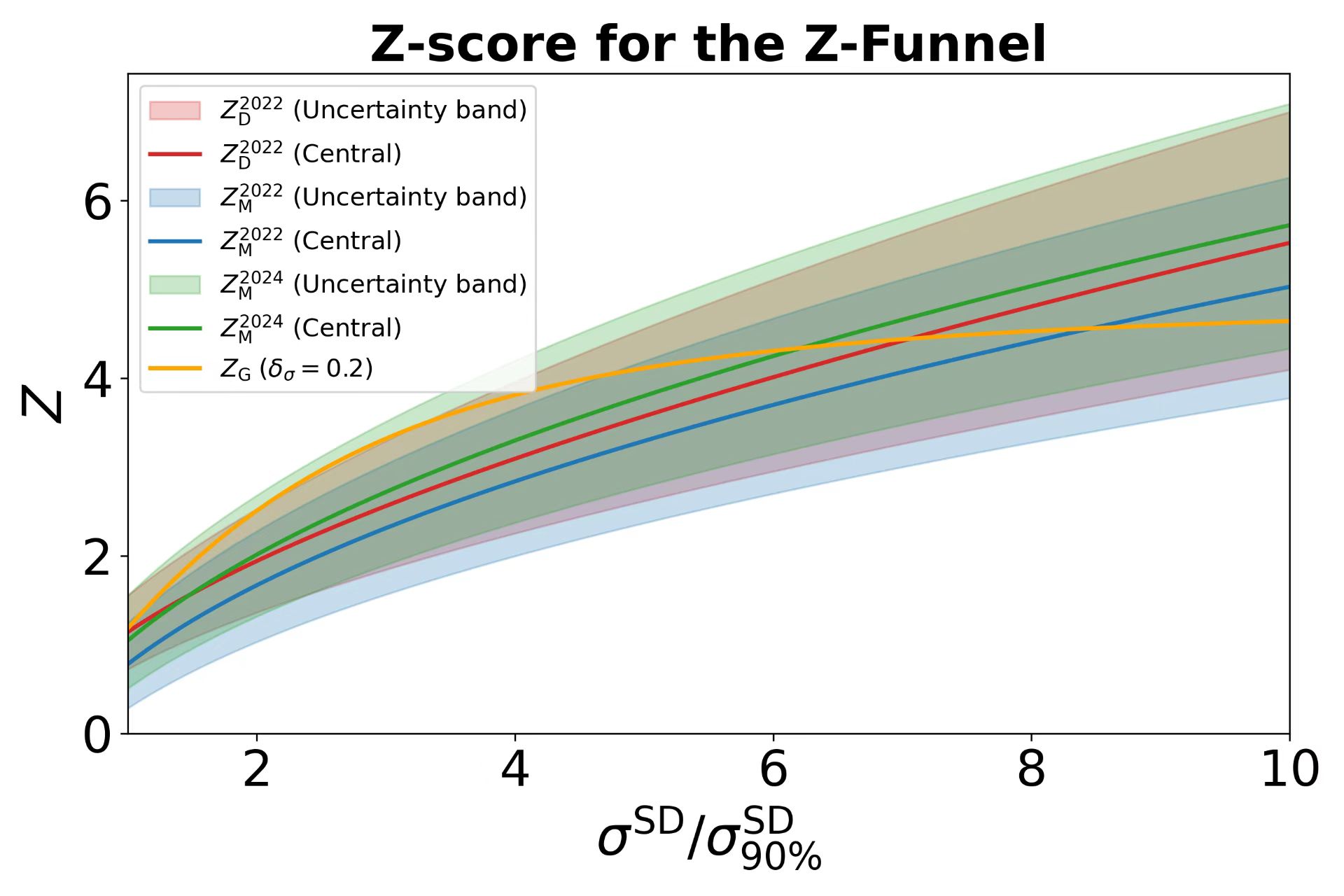}\\
	\caption{One-sided Gaussian-equivalent significance $Z$ as a function of $\sigma^{\rm SI}/\sigma^{\rm SI}_{90\%}$ (left) and $\sigma^{\rm SD}/\sigma^{\rm SD}_{90\%}$ (right), obtained with \texttt{DDCalc} and \texttt{micrOMEGAs} for \textbf{LZ-2022}, with \texttt{micrOMEGAs} for \textbf{LZ-2024}, and with a simplified Gaussian likelihood. Solid curves use the central cross sections; shaded bands indicate the variation from the hadronic uncertainties specified in the text. At the point $\sigma/\sigma_{90\%}=1$, $Z_{\rm D}^{2022}= 1.16~(1.14)$, $Z_{\rm M}^{2022} = 1.03~(0.78)$, and $Z_{\rm M}^{2024} = 1.09~(1.04)$ for $h$ ($Z$) funnel, all below $Z_{90\%} = 1.28 $ for $90\%$ C. L. exclusion. These results reflect that \texttt{DDCalc} provides a more realistic recast of the LZ experiment, albeit both methods are conservative in estimating the exclusion significance. }
	\label{Chi2plots}
\end{figure}

\subsection*{Comparison of $Z$-score calculations}

For the \textbf{LZ-2022} data, \texttt{DDCalc} constructs the binned Poisson likelihood
\begin{equation}
\mathcal L_{\rm DDC}=\prod_i
\frac{(\theta\, b_i+N_{{\rm ev},i})^{N_{{\rm obs},i}}\,e^{-(\theta\, b_i+N_{{\rm ev},i})}}{N_{{\rm obs},i}!}
\times\mathcal G(\theta),
\label{eq:LDDC}
\end{equation}
where $N_{{\rm obs},i}$, $b_i$, and $N_{{\rm ev},i}$ are the observed event count, the expected background, and the expected signal in the $i$th analysis bin, respectively, and $\mathcal G(\theta)$ is a Gaussian constraint on the background normalization $\theta$, which is profiled~\cite{GAMBITDarkMatterWorkgroup:2017fax, GAMBIT:2018eea}. The test statistic $q$ and its conversion to the one-sided significance $Z_{\rm D}$ are defined in the main text. The \texttt{micrOMEGAs} recast instead assumes a single bin with no background and no observed events ($b=0$, $N_{\rm obs}=0$), so that Eq.~\eqref{eq:LDDC} reduces to $\mathcal L_{\rm M}=e^{-N_{\rm ev}}$, which coincides with the $p$-value.

Fig.~\ref{Chi2plots} shows the resulting significances for SI (left) and SD (right) scattering as functions of $r\equiv\sigma^{\rm SI(SD)}/\sigma^{\rm SI(SD)}_{90\%}$, where $\sigma^{\rm SI(SD)}_{90\%}$ is the published 90\%~C.L.\ upper limit of the respective data set. Normalizing to each limit makes the curves for different data sets directly comparable. We denote the \texttt{DDCalc} and \texttt{micrOMEGAs} results for \textbf{LZ-2022} by $Z_{\rm D}^{2022}$ (red) and $Z_{\rm M}^{2022}$ (green), and the \texttt{micrOMEGAs} result for \textbf{LZ-2024} by $Z_{\rm M}^{2024}$ (blue). Because the predicted cross sections in the funnel regions exceed the \textbf{LZ-2024} limits by up to an order of magnitude, we show $1\le r\le10$. Solid curves correspond to the central cross sections; the shaded bands span the hadronic uncertainties, with upper and lower edges corresponding to shifts of $+25\%$ ($+47\%$) and $-22\%$ ($-38\%$) in $\sigma^{\rm SI}$ ($\sigma^{\rm SD}_n$).

For comparison, we also show the significance obtained from a simplified Gaussian likelihood (orange),
\begin{equation}
Z_{\rm G}^{\rm }=\frac{\sigma}{\sqrt{(\sigma_{90\%}/1.64)^2+(\delta_\sigma\sigma)^2}},
\label{eq:ZG}
\end{equation}
with a fixed relative uncertainty $\delta_\sigma=0.20$. Although Eq.~\eqref{eq:ZG} resembles the objective function used for sampling (cf. Eq.~(\ref{eq:gaussian})), it serves here only as a proxy for the exclusion significance. Since $Z_{\rm G}$ depends on $\sigma$ only through $r$, it is the same for both data sets.


Fig.~\ref{Chi2plots} shows the following features:
\begin{itemize}
\item $Z_{\rm M}$ and $Z_{\rm D}$ approximately coincides at $r=1$ by construction, since they are calibrated to the published $90\%$~C.L. limit, and they remain close for $\sigma/\sigma_{90\%}\lesssim2$; moderate differences develop at larger ratios, where the two statistical constructions differ. Most importantly, $Z_{\rm M}^{2022} <Z_{\rm D}^{2022}$ over the entire range, confirming that the background-free Poisson recast is conservative relative to the full likelihood analysis of the same data. Because the \texttt{micrOMEGAs} treatment of \textbf{LZ-2024} follows the identical construction, we expect $Z_{\rm M}^{2024}$ to be likewise conservative with respect to a dedicated likelihood analysis of the \textbf{LZ-2024} data, which is not possible at present owing to the limited public information. This speculation is consistent with the observation that the \texttt{micrOMEGAs} reference cross section $\sigma^{\rm SI}_0$ for \textbf{LZ-2024} lies above the official limit in the funnel mass range, as illustrated by Fig.~3 of Ref.~\cite{Belanger:2026asz}.
    
\item $Z_{\rm G}$ follows the qualitative trend of both methods but overestimates the significance for $1\le r\lesssim4$--$5$. Note that the curve labeled $Z_{\rm G}$ in Fig.~\ref{Chi2plots} is obtained from the exact one-sided formula $Z_{\rm G} = \sqrt{2} \; \text{Erfc}^{-1}[2\; \text{Erfc}(\sqrt{\chi_{\rm G}^2/2})]$, with $\chi_{\rm G}^2 = -2\log\left(\mathcal{L}_{\rm DD}\right)$, which gives $Z_{\rm G}=1.18$ at $r=1$. 
At larger $r$, $Z_{\rm G}$ saturates and falls below $Z_{\rm M}$ and $Z_{\rm D}$, since an uncertainty proportional to $\sigma$ bounds it from above by $1/\delta_\sigma=5$. Gaussian approximations of this type are thus unsuitable for assessing high-significance exclusions.
\end{itemize}

\begin{figure}[]
	\centering
	\includegraphics[width=0.44\linewidth]{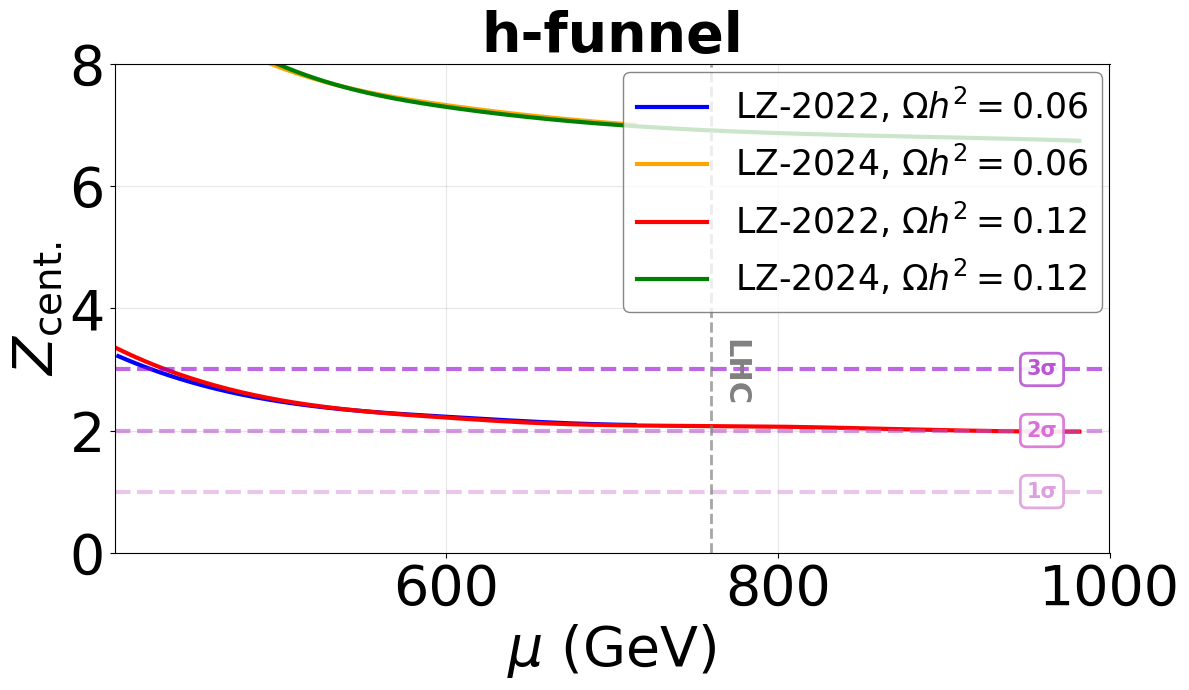}
	\includegraphics[width=0.44\linewidth]{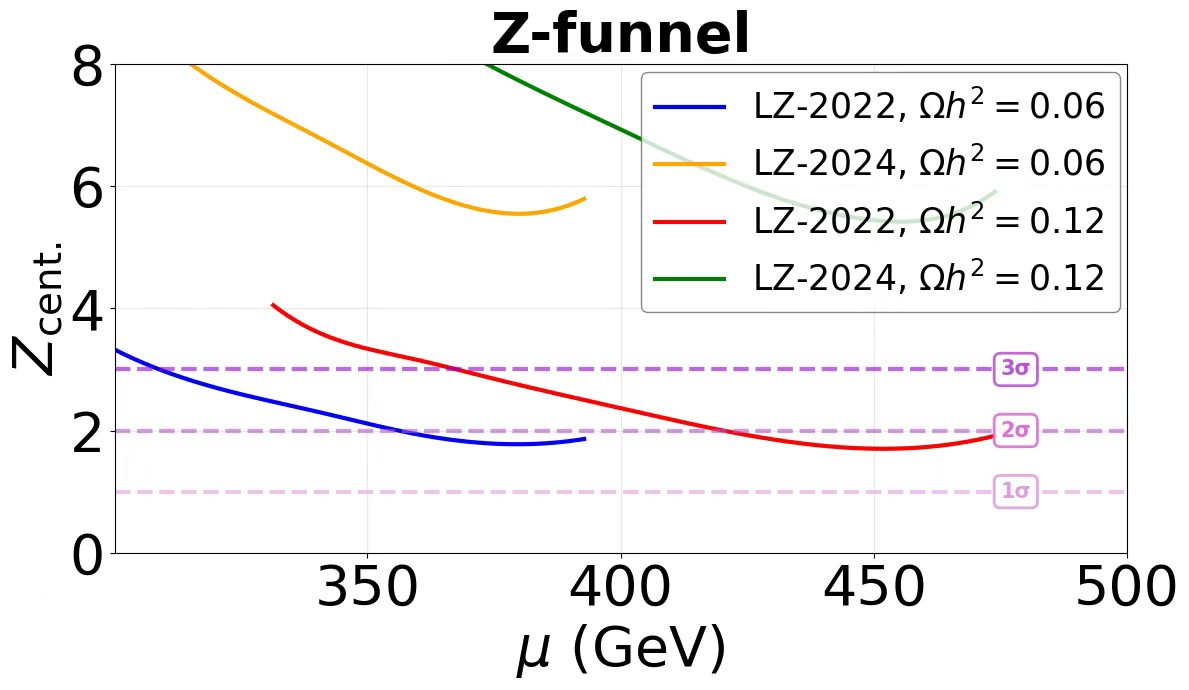}\\
	\caption{Same as Fig.~\ref{fig:chi2minhzfunnel} except that the scattering cross-sections are calculated with the central values of the nucleon matrix elements.}
	\label{Comparison}
\end{figure}

\subsection*{Scanning strategy}

The parameter space of Table~\ref{tab:parameters} is explored with the nested-sampling algorithm \texttt{MultiNest}~\cite{Feroz:2008xx}, guided by the objective function of Eq.~(\ref{eq:gaussian}). Because this function serves only to concentrate the samples near the boundary of the allowed region, the quantities presented in the main text are envelopes of the sampled points rather than credible regions. The following procedure ensures that these envelopes are determined reliably.

\paragraph{Partitioning of the scans.}
We consider two representative relic densities, $\Omega h^2=0.12$ and $0.06$, for which $\widetilde\chi^0_1$ constitutes all or half of the DM, respectively; in the latter case the expected event rates in LZ are rescaled by $\Omega h^2/0.12$. For each relic density and each funnel, the parameter space is divided into four regions according to $\mathrm{sign}(M_1)$ and $\mathrm{sign}(A_t)$, and each region is scanned independently with $n_{\rm live}=10^4$ live points~\cite{Feroz:2008xx}, giving $2\times2\times4=16$ scans in total. This partitioning prevents the sampler from settling on a single sign combination and ensures that regions with suppressed $h$- or $Z$-mediated contributions, such as the vicinity of the blind spot at $M_1\mu<0$, are sampled with comparable density.

\paragraph{Stability against the objective function.}
The widths $\delta_{\Omega h^2}$ and $\delta_\sigma^{\rm SI}$ in Eq.~(\ref{eq:gaussian}) control only how strongly the sampler is focused on the target relic density and on small cross sections, and should not affect the envelopes. We verify this for $\Omega h^2=0.12$ in the regions that yield the lowest cross sections, namely $M_1>0$, $A_t>0$ for the $h$ funnel and $M_1<0$, $A_t>0$ for the $Z$ funnel, by repeating the scans with $\delta_{\Omega h^2}=5\%,~10\%,~20\%$ and $\delta^{\rm SI}_\sigma=\delta^{\rm SD}_\sigma=10\%,~20\%,~30\%$. The lower envelopes of $\sigma^{\rm SI}$ and $\sigma^{\rm SD}_n$ agree in all cases.

\paragraph{Cross-check with a genetic algorithm.}
To test independently whether the minimal cross sections have been found, we fix $\Omega h^2=0.12$ and minimize $\sigma^{\rm SI}$ in the $h$ funnel and $\sigma^{\rm SD}_n$ in the $Z$ funnel with the genetic algorithm \texttt{PIKAIA}~\cite{Charbonneau:1995}. The minima agree with the lower envelopes obtained with \texttt{MultiNest}. The genetic algorithm converges faster but returns only the minimum rather than the distribution of points; we therefore use \texttt{MultiNest} for the results in the main text.

\paragraph{Dependence on the hadronic treatment.}
Fig.~\ref{Comparison} shows the minimal significance $Z_{\min}$ as a function of $\mu$, computed with the central values of the hadronic matrix elements. For \textbf{LZ-2024}, $Z_{\min}$ significantly exceeds the values in Fig.~\ref{fig:chi2minhzfunnel}, where the cross sections are rescaled by the adverse factors $0.78$ (SI) and $0.62$ (SD), over the whole range of $\mu$. The significances quoted in the main text are therefore conservative with respect to the hadronic uncertainties.

\end{document}